\documentclass[reprint, amsmath,amssymb, aps,superscriptaddress,longbibliography]{revtex4-1}

\usepackage{graphics}
\usepackage{dcolumn}
\usepackage{bm}
\usepackage[caption=false]{subfig}
\usepackage{xfrac}
\usepackage{mathrsfs}
\usepackage{float}
\usepackage[normalem]{ulem}
\usepackage[bbgreekl]{mathbbol}
\usepackage[utf8]{inputenc}
\usepackage{amsmath} 
\usepackage{multirow}
\usepackage{adjustbox}
\usepackage{etoolbox}
\usepackage{comment}
\usepackage{amsmath}
\usepackage{amssymb}
\usepackage{blindtext}
\usepackage{hyperref}
\usepackage{color}
\newcommand{\figref}[1]{Fig.~\ref{fig:#1}}
\newcommand{\beginfigref}[1]{Figure~\ref{fig:#1}}
\newcommand{\secref}[1]{Sec.~\ref{#1}}
\newcommand{\begineqref}[1]{Equation~(\ref{eq:#1})}
\renewcommand{\eqref}[1]{Eq.~(\ref{eq:#1})}
\newcommand{\eqsref}[2]{Eq.~(\ref{eq:#1}) and Eq.~(\ref{eq:#2})}
\newcommand{\eqssref}[2]{Eqs.~(\ref{eq:#1}-\ref{eq:#2})}
\renewcommand{\vec}[1]{\mathbf{#1}}

\begin{document}

\title{Polarization dependence of the propagation constant of leaky guided modes}

\author{Adi Pick}
\email{adipick@technion.ac.il}
\affiliation{Faculty of Chemistry, Technion-Israel Institute of Technology, Haifa 32000, Israel.}
\affiliation{Faculty of Electrical Engineering, Technion-Israel Institute of Technology, Haifa 32000, Israel.}
\author{Nimrod Moiseyev}
\affiliation{Faculty of Chemistry, Technion-Israel Institute of Technology, Haifa 32000, Israel.}
\affiliation{Faculty of Physics, Technion-Israel Institute of Technology, Haifa 32000, Israel.}

\begin{abstract}
We show that transverse-magnetic (TM) leaky modes can propagate further than transverse electric (TE) modes in real-index dielectric waveguides. We compute the density of states and find that while the TE spectrum contains only overlapping resonances, the TM spectrum typically contains several isolated peaks. By transforming the TM equation into a Schr\"{o}dinger-type equation, we show that these isolated peaks arise due to $\delta$-function barriers at the core-cladding interface. Our theory is useful for a range of applications, including filtering TM modes from initially unpolarized light and transferring information between distant waveguides.
\end{abstract}

\maketitle
\section{Introduction}

The  analogy between Maxwell's equations for  light propagation in lossy   waveguides and     non-Hermitian   quantum mechanics~\cite{bender1998real,ruter2010observation,klaiman2008visualization,makris2008beam}  has  lead  to the discovery of many  intriguing phenomena, such as loss-induced transparency~\cite{Guo2009},  gain-induced suppression of lasing~\cite{brandstetter2014reversing},  unidirectional invisibility~\cite{lin2011unidirectional},  adiabatic optical  switches~\cite{vorobeichik1998intermediate,doppler2016dynamically}, and sensors with sub-linear sensitivity~\cite{wiersig2014enhancing}. In this work, we report yet another intriguing property of  non-Hermitian waveguides, which stems from  the analogy to quantum mechanics: Transverse-magnetic (TM) leaky modes  can propagate further than transverse electric (TE) modes along real-index thin waveguides, and are  more suitable  for applications which require isolated resonances.  
  
In the most simple picture, an optical fiber consists of a high-index material (core) coated by a lower-index material (cladding)~\cite{snyder2012optical}.  In the absence of    loss or gain, light at certain frequencies and wavelengths  is confined  to propagate  inside the core due to total internal reflection at the core-cladding interface~\cite{Jannopoulos2008}.  These   are the so-called  \emph{confined guided modes}, which  propagate   along the fiber while  accumulating  an overall phase of $e^{i\beta_n z}$ with a real propagation constant $\beta_n$. However, in the presence of material absorption, radiation loss, or gain, light can be  attenuated or amplified upon propagation. In such cases, the propagation constant $\beta_n$ is complex~\cite{skorobogatiy2009fundamentals}, and the modes are called \emph{leaky guided modes}~\cite{hu2009understanding}.   
When  the light  intensity is attenuated along the propagation   direction, it grows unboundedly  in the transverse  direction [as follows from the dispersion relation, \eqref{wave-relations-a}]. This divergence poses many  theoretical challenges, such as finding a proper way to  normalize the modes~\cite{leung1994completeness,lee1999dyadic,pick2017general} and revisiting various  expressions from  ``Hermitian optics''~\cite{Lee2008,kristensen2012generalized}.   
While most previous work on  complex-propagation constants typically involves  gain or loss in the waveguide~\cite{marcuvitz1951waveguide,novotny1994light,schlereth1990complex,anemogiannis1992multilayer}, we explore  in this work the less familiar   case, where   $\beta_n$ is complex solely due to radiation losses in the transverse direction~\cite{vorobeichik1995modal}.  In the latter type of modes, $\beta_n$ strongly depends on the polarization and, consequently, the polarization can be used as a knob to control the propagation.

Despite the long-standing debate on the interpretation, completeness,   and normalization of leaky modes~\cite{leung1994completeness,lee1999dyadic},  there is no question about their usefulness  when it comes to  describing light at nearly resonant wave vectors and in close proximity to the waveguides.  Most importantly,  the  complex propagation constants $\beta_n$ determine the location of peaks in the density states.  This is  similar to non-Hermitian  quantum mechanics, where  resonant complex eigenenergies, $E_n = \varepsilon_n - i\Gamma_n$, represent peaks in the density of  continuum states, centered around  real energies $\varepsilon_n$ with width $\Gamma_n$~\cite{Moiseyev2011}.  In this work, we use the term \emph{isolated resonances } when the peaks   do not overlap  (or, more  formally, when $|\varepsilon_{n+1} - \varepsilon_n| >\Gamma_n,\Gamma_{n+1}$).

\begin{figure*}[t]
                 \includegraphics[width=1\textwidth]{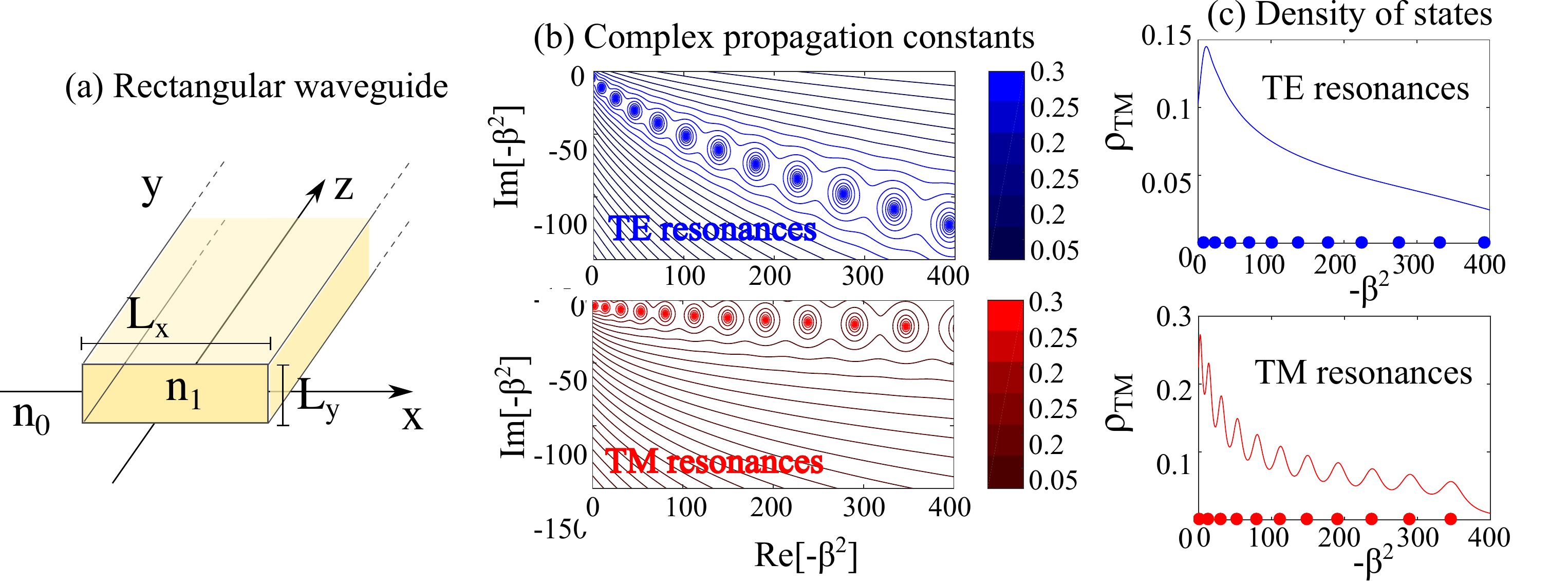}
                 \centering
                                   \caption{(a) Dielectric waveguide with a thin rectangular cross section ($L_x\gg L_y$ and  $L_x = 1\mu m$) and   index $n_1 = \sqrt{2}$   surrounded by a medium with index $n_0 = 1$ (assuming    $n_1>n_0\geq1$).     The  wavelength of light is $\lambda =\tfrac{2\pi}{\omega}= 3\mu m$.  (b)  TE and TM complex propagation constants for the structure from (a). The plots depict   contours  of the functions  $\Delta_\mathrm{TE}(\beta^2)$ (top) and $\Delta_\mathrm{TM}(\beta^2)$ (bottom) (defined in  text), whose poles give the resonant modes.      (c) Density of states,  evaluated using \eqref{DOS}, for TE   (top) and  TM (bottom) modes. The dots mark the  real parts of  propagation constants  from (b). 
}
\label{fig:general-idea}
  \end{figure*}

\beginfigref{general-idea} summarizes the main result of this paper: the existence of   narrow TM  resonances in real-index dielectric waveguides. 
 We consider here the rectangular waveguide  shown  in \figref{general-idea}(a).  Since the system has mirror-plane symmetry around $z=0$, the  waveguide can support either TE or TM modes, in which    the electric or  magnetic fields are  transverse to the direction of propagation. In \secref{scalar-maxwell}, we review the scalar Maxwell equations for TE and TM polarizations [\eqref{TE} and \eqref{TM} respectively] and  in \secref{ME-solutions}, we present their solution, which demonstrates  the polarization dependence  of the propagation constants. \beginfigref{general-idea}(b) shows contour plots of  the  solutions of the transcendental equations from \secref{ME-solutions}  [\eqssref{cond_a}{cond_d}], whose zeros are the TE and TM resonant  propagation constants. { In electromagnetic scattering theory, these resonant propagation constants are the  infinite eigenvalues of the scattering matrix~\cite{newton2013scattering}.}  Clearly,  the TM resonances  are  situated  closer to the real axis and are, therefore,    more strongly confined to the waveguide. We explain this result in \secref{transformation}, by  using the analogy between Maxwell's equations and the  Schr\"{o}dinger equation.  In  \secref{DOS-section}, we explore an important consequence  of  the  narrow TM resonances: the appearance of isolated peaks in the TM density of states [as shown in \figref{general-idea}(c)].   Having a proper understanding of the polarization dependence of leaky modes in thin waveguides, the next step is to design simple structures with narrow TM resonances, which will benefit from this effect.  In \secref{discussion-section}, we describe two possible applications of our theory for filtering TM modes from initially unpolarized light  and for  transferring  information  between distant waveguides. 
 
\section{Scalar Maxwell  equations\label{scalar-maxwell}}

Our starting point is   the  frequency-domain Maxwell equations
for nonmagnetic media~\cite{Jannopoulos2008}:  $\nabla\times \vec{E} = i\omega \mu_0\vec{H}$, and $\nabla\times\vec{H}= -i\omega \varepsilon_0\varepsilon\vec{E}$. Here, $\vec{E}$ and $\vec{H}$ are the electric and magnetic vector fields, $\varepsilon_0$ and $\mu_0$ are the vacuum permittivity and permeability, and $\varepsilon$  is the relative permittivity of the medium (the relative permeability is 1). From Maxwell's equations, one  can  easily obtain two decoupled   wave equations for the electric and magnetic fields~\cite{Jannopoulos2008}:
\begin{gather}
\nabla\times\nabla\times \vec{E} = \left( \tfrac{\omega}{c}\right)^2\varepsilon\, \vec{E} \label{eq:Maxwell-wave2}\\
\nabla\times\tfrac{1}{\varepsilon}\nabla\times \vec{H} =  \left( \tfrac{\omega}{c}\right)^2\vec{H}, 
\label{eq:Maxwell-wave1}
\end{gather}
with the speed of light given by $c = 1\sqrt{\varepsilon_0\mu_0}$.  Due to the symmetry of the geometry under study [\figref{general-idea}(a)], the eigenmodes are either transverse electric (with non-zero field components   $E_y$, $H_x$, and $H_z$) or transverse magnetic (with non-zero  $H_y$, $E_x$ and $E_z$). This property allows to reduce Maxwell's vectorial  equations [\eqsref{Maxwell-wave2}{Maxwell-wave1}] to  scalar equations for the electric and magnetic fields.  

 Since high aspect-ratio waveguides are known to have  record-low losses \cite{bauters2011ultra,dai2010polarization}, we consider   in this work    thin rectangular waveguides, as shown  in~\figref{general-idea}(a).
In this limit ($L_x\gg L_y$),  the $y$-dependence of the field can be neglected. The electric and magnetic modes  have  the form
\begin{equation}
{\psi}(x,z) = e^{i\beta z}{\psi}(x),
\label{eq:guided}
\end{equation}
and the propagation constant $\beta$ is generally complex. 
Focusing first on   TE polarization,  we substitute $E_y = e^{i\beta z}e_y(x)\hat{y}$ into \eqref{Maxwell-wave2},   introduce the index of refraction $n^2 =\varepsilon$, and obtain
\begin{align}
\left[\frac{d^2}{dx^2} +  \left( \tfrac{\omega}{c}\right)^2n^2(x)\right]e_y(x) = \beta^2 e_y(x).
\label{eq:TE}
\end{align}
This equation  has  precisely the same form as the time-independent Schr\"{o}dinger equation of a one-dimensional particle:
\begin{align}
&\left[-\frac{\hbar^2}{2m}\frac{d^2}{dx^2} +V(x)\right]\psi(x) = E \psi(x),
\label{eq:SE-well}
\end{align}
 with  $m = 0.5$, $\hbar = 1$,   $E = -\beta^2$ and potential field 
\begin{equation}
V_\mathrm{TE}(x) = - \left( \tfrac{\omega}{c}\right)^2 n^2(x).
\label{eq:V-TE}
\end{equation}
The situation is quite different for the TM polarization.  Substituting $H_y =  e^{i\beta z}h_y(x)\hat{y}$ into \eqref{Maxwell-wave1}, one finds that the magnetic field satisfies the scalar equation
\begin{equation}
-\tfrac{d}{dx}\tfrac{1}{\varepsilon}\tfrac{d}{dx}h_y + \beta^2\tfrac{1}{\varepsilon}h_y =  \left( \tfrac{\omega}{c}\right)^2h_y,
\label{eq:TM-equation-BCs}
\end{equation}
or alternatively
\begin{align}
\left[\frac{d^2}{dx^2} +  \left( \tfrac{\omega}{c}\right)^2n^2(x) - \frac{d\ln n^2(x)}{dx}\frac{d}{dx}\right]h_y(x) = \beta^2 h_y(x)
\label{eq:TM}
\end{align}
(For details on how to obtain this result, see~\cite{snyder2012optical} or  \footnote{
We use the identity 
$\varepsilon\tfrac{d}{dx}\tfrac{1}{\varepsilon}\tfrac{d}{dx}h_y = 
-\tfrac{\varepsilon'}{\varepsilon}\tfrac{dh_y}{dx} + \tfrac{d^2h_y}{dx^2} = -\tfrac{d \ln n^2}{dx}\tfrac{dh_y}{dx} + \tfrac{d^2h_y}{dx^2}$.}.)
The last term in square brackets contains a spatial derivative and, therefore, cannot be interpreted as the  potential field of a conservative force. In \secref{transformation}, we will transform  \eqref{TM} into an equivalent Schr\"{o}dinger-type equation with an effective conservative potential, and show that   this term is responsible for the narrow TM resonances.

\section{Confined and leaky modes\label{ME-solutions}}
Our example system from \figref{general-idea}(a) can be solved semi-analytically using standard techniques from quantum mechanics~\cite{gasiorowicz2007quantum}.  The  eigenmodes of a piecewise homogeneous potential   are outgoing  planewave solutions, whose coefficients are determined by matching the field and  its derivatives at the boundaries.  Since our example problem  is symmetric under reflection around $x=0$, it is convenient to use the ansatz:
\begin{equation}
  \psi(x)=\begin{cases}
  e^{iq x}, & \text{for $x<-\frac{L}{2}$}\\
    A\cos(k_xx) + B\sin(k_xx), & \text{for $|x|<\frac{L}{2}$}\\
        e^{-iq x}, & \text{for $x>\frac{L}{2}$},
  \end{cases}
  \label{eq:ansatz}
\end{equation}
where even and  odd solutions have  $B=0$ and $A=0$ respectively. Here,  $\psi$  is either $E_y$ (for TE modes) or $H_y$  (for TM modes) and 
the $x$-components of the wave vectors in the core and cladding, $k_x$ and $q$,  are related to the propagation constant, $\beta$, via the  dispersion relations
\begin{gather}
k_x^2 + \beta^2 = \left(\tfrac{\omega}{c}\right)^2\varepsilon_1
\label{eq:wave-relations-a}\\
q^2 + \beta^2 = \left(\tfrac{\omega}{c}\right)^2\varepsilon_0.
\label{eq:wave-relations-b}
\end{gather}

\begin{figure}
 \centering
 \includegraphics[width=\linewidth]{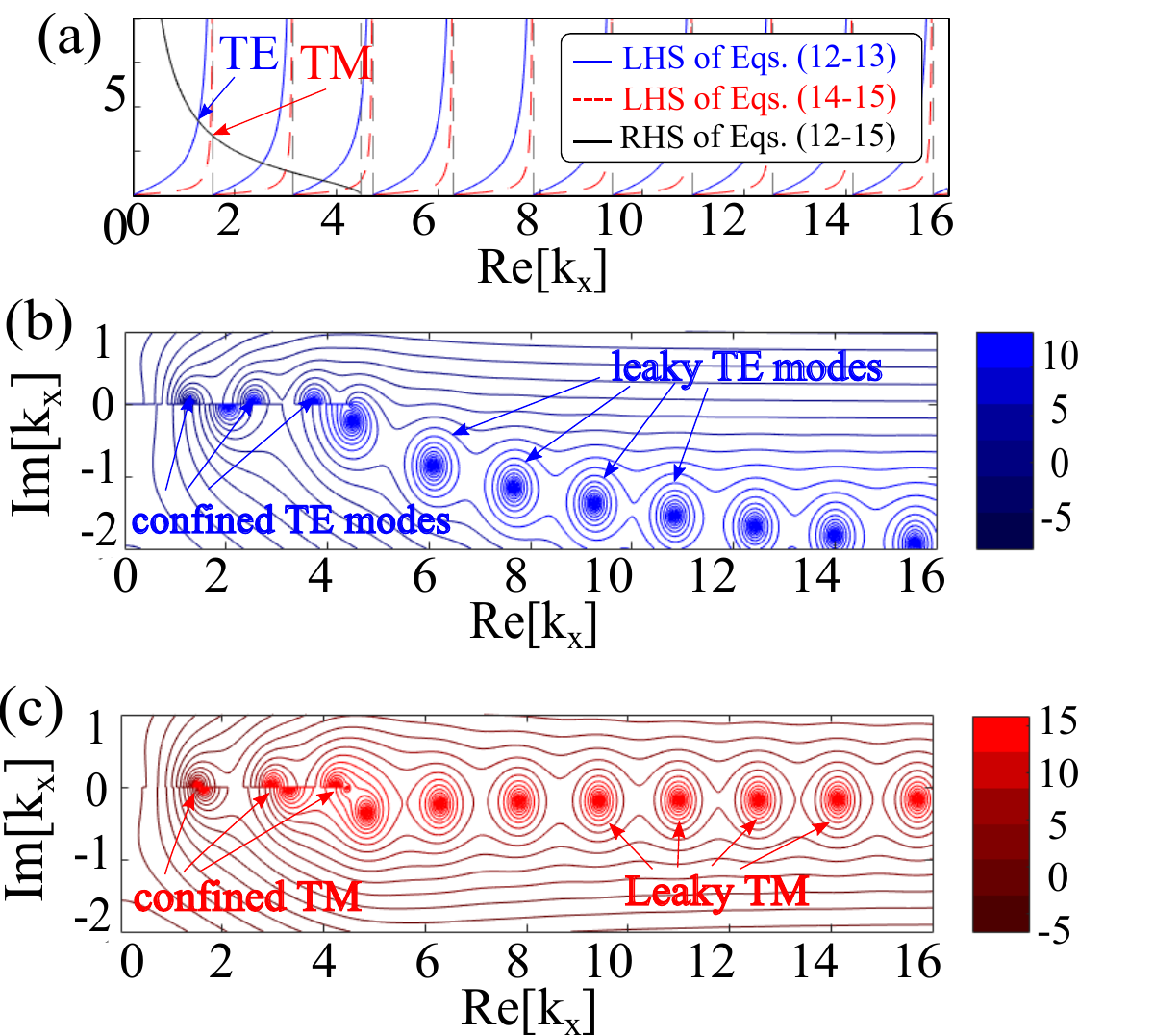}
 \caption{	(a) Right and left-hand sides of the transcendental equations  \eqssref{cond_a}{cond_d} for the structure from \figref{general-idea}(a) (with $n_0$ = 1, $n_1 = \sqrt{2}$, $\omega = 2$). The intersections of the  blue-solid (red-dashed) curves  with the black curve define the  transverse wave vectors $[k_x^\mathrm{con}]_n$ of  TE (TM) confined guided modes. (b) and (c) show contours of the functions $\Delta_\mathrm{TE}(k_x)$ and $\Delta_\mathrm{TM}(k_x)$ respectively (defined in the text), whose complex poles are the transverse wave vectors of confined and leaky TE/TM modes. }
 \label{fig:transcendental contours}
\end{figure}

Since the TE equation [\eqref{TE}]  is equivalent to a   one-dimensional particle in a box, the boundary conditions   are   continuity of the field ($\psi$)   and its  derivative  ($d\psi/dx$) at  the core-cladding interface ($x=\pm L/2$). By demanding continuity of $\psi$ and $d\psi/dx$ for the ansatz solution  [\eqref{ansatz}]  and  using the dispersion relations [\eqsref{wave-relations-a}{wave-relations-b}] to   express $q$ in terms of $k_x$, one obtains the well-known  transcendental equations~\cite{snyder2012optical}: 
\begin{align}  
& \mathrm{Even\,\, TE \,\,modes:\,}\tan(\tfrac{k_xL}{2}) = -i\sqrt{1 - \tfrac{\omega^2(\varepsilon_1 - \varepsilon_0)}{(ck_x)^2}}\label{eq:cond_a}\\ 
 &\mathrm{Odd\,\, TE \,\,modes:\,}-\cot(\tfrac{k_xL}{2}) = -i\sqrt{1 - \tfrac{\omega^2(\varepsilon_1 - \varepsilon_0)}{(ck_x)^2}}\label{eq:cond_b}
\end{align}


   \begin{figure*}[t]
   \centering
     \includegraphics[scale=0.6]{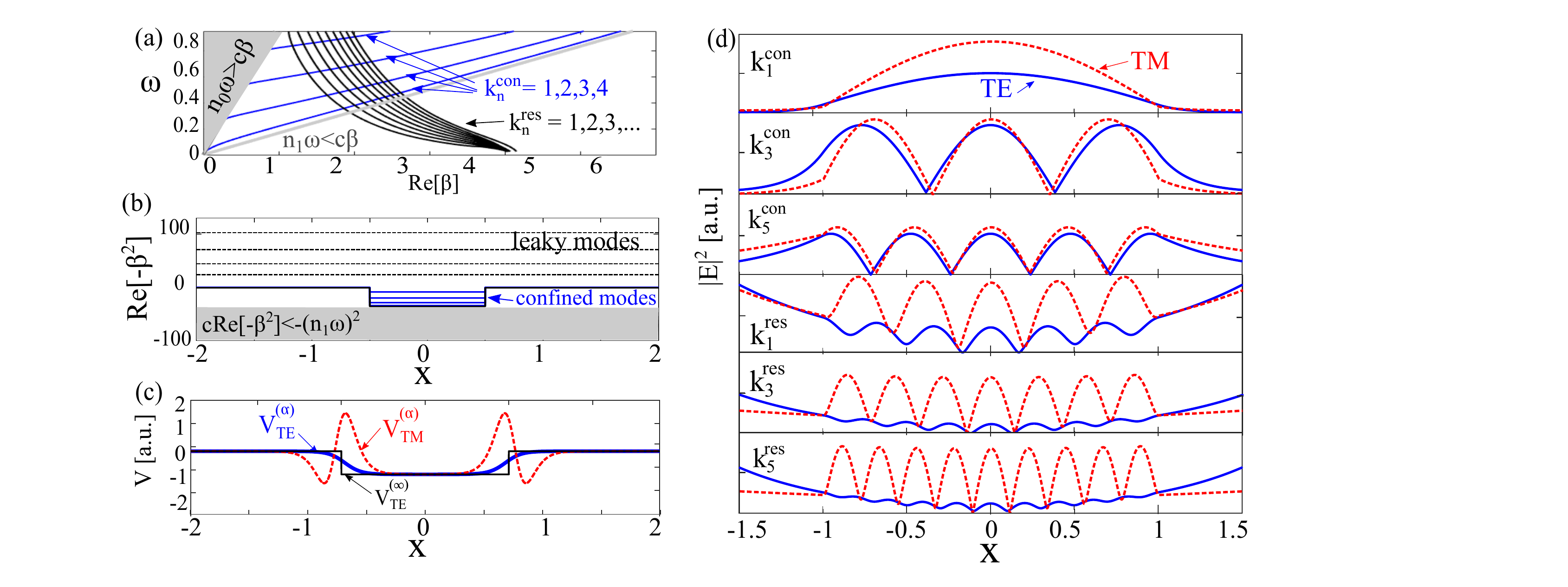}
     \caption{(Color online) (a) Dispersion relation ($\omega$ vs. $\mathrm{Re}[\beta_n]$) of TE modes,
obtained by computing  confined (blue) and leaky (black)  propagation constants, $\beta_n$, at a range of frequencies, $0.01<\tfrac{\omega L}{2\pi c}<1$, for the structure from \figref{general-idea}. Confined modes   propagate in the core and decay in the cladding and satisfy $\omega\varepsilon_0<\beta_nc<\omega\varepsilon_1$. The gray shaded area marks the light line of the cladding ($\beta_nc>\omega n_0$).  (b) Confined (blue solid) and leaky (black dashed)  TE  modes and the TE potential  (black solid)  [\eqref{V-TE} with $n(x)$ given by \eqref{sharp-n}], with $\tfrac{\omega L}{2\pi c} = 2$.
    (c) Smoothed TE (blue solid) and TM (red  dotted) potentials, obtained by evaluating \eqsref{V-TE}{potential} respectively, using the smoothed index profile \eqref{smooth-n}]. { (d) TE and TM confined and leaky mode profiles  (blue solid and red  dotted lines respectively).}  
  }
    \label{fig:dispersion-fig}
  \end{figure*}

In contrast, the TM equation [\eqref{TM}] contains an additional derivative term which changes the boundary conditions. In order to derive the correct boundary conditions, one  can  integrate \eqref{TM-equation-BCs}  over an infinitesimal region around the boundary (at $x = \tfrac{L}{2}$).  The first term on the left-hand side  gives
\begin{gather*}
\lim_{\delta\rightarrow0} 
\int_{\tfrac{L}{2}-\delta}^{\tfrac{L}{2}+\delta} 
 dx\,
\tfrac{d}{dx}\tfrac{1}{\varepsilon} \tfrac{d h_y }{dx} =    
\tfrac{h_y '({L}/{2})_\mathrm{out}}{\varepsilon(L/2)_\mathrm{out}} -  \tfrac{h_y '(L/2)_\mathrm{in}}{\varepsilon(L/2)_\mathrm{in}}
\end{gather*}
and the remaining terms vanish.
Therefore,   the TM transcendental equations are~\cite{snyder2012optical}:
\begin{align}  
& \mathrm{Even\, TM \,modes:\,}\tfrac{\varepsilon_0}{\varepsilon_1}\tan(\tfrac{k_xL}{2}) = -i\sqrt{1 - \tfrac{\omega^2(\varepsilon_1 - \varepsilon_0)}{(ck_x)^2}}\label{eq:cond_c}\\ 
 &\mathrm{Odd\, TM \,modes:\,}\tfrac{\varepsilon_0}{\varepsilon_1}-\cot(\tfrac{k_xL}{2}) = -i\sqrt{1 - \tfrac{\omega^2(\varepsilon_1 - \varepsilon_0)}{(ck_x)^2}}\label{eq:cond_d}.
\end{align}

\beginfigref{transcendental contours}(a)  shows  the  TE and TM  \emph{confined guided modes} for the structure from \figref{general-idea}(a), which correspond to  real-$k_x$  solutions of   \eqssref{cond_a}{cond_d}. Graphically, real-$k_x$ solutions are found by intersecting  the  blue (TE) and red (TM) curves [the left-hand sides of  \eqssref{cond_a}{cond_b} and \eqssref{cond_c}{cond_d} respectively] with  the  black  curve [the right-hand side of \eqssref{cond_a}{cond_d}]. Since the TE and TM equations only differ in the factor   $\tfrac{\varepsilon_0}{\varepsilon_1}$, which determines the slope of the tangent and cotangent  functions but not  the location of the branch cuts, the number of TE and TM confined  modes is the same for any given index contrast, but  TM modes are shifted to larger $k_x$ values.

Panels (b-c) in \figref{transcendental contours} show, in addition to the confined modes,  the  TE and TM  \emph{leaky guided modes}, which correspond to  complex-$k_x$ solutions of \eqssref{cond_a}{cond_d}. It is evident from the figure that   the TM resonances are closer to the real axis in comparison to the TE resonances, which implies that a larger fraction of the TM-modal intensity  is  confined in the core of the waveguide. Formally, the even/odd modes are given by  the  zeros of the functions $\mathcal{F}_\mathrm{TE}^{(e/o)}$ and  $\mathcal{F}_\mathrm{TM}^{(e/o)}$,  defined as the difference between the left- and right-hand sides of \eqssref{cond_a}{cond_d}. \beginfigref{transcendental contours} shows the poles of $\Delta_\mathrm{TE} \equiv |\mathcal{F}_\mathrm{TE}^{(e)}|^{-2} + |\mathcal{F}_\mathrm{TE}^{(o)}|^{-2}$ and $\Delta_\mathrm{TM} \equiv |\mathcal{F}_\mathrm{TM}^{(e)}|^{-2} + |\mathcal{F}_\mathrm{TM}^{(o)}|^{-2}$. 
{ These poles  are precisely the well-known scattering matrix poles, which can be derived directly from Maxwell’s equations using electromagnetic scattering theory~\cite{newton2013scattering}. The location of the poles in the complex plane determines many physical properties, such as the scattering, absorption, and extinction cross sections. }
Note that  despite the fact that we expect, based on \figref{transcendental contours}(a), to find three  real-$k_x$ solutions both in the TE and TM polarizations, panels (b) and (c) show   spurious real-$k_x$ solutions [e.g., the pole on the real axis in (b) at $k_x\approx2$]. 
These additional poles are an artifact of our numerical procedure, since we  plot contours of the inverse squared modulus   of the boundary-condition equations and not the equations themselves.

We conclude this section by discussing the dispersion relation of the guided modes, presented in \figref{dispersion-fig}a.  Confined guided modes    propagate inside the core and decay in the cladding. Since these  modes have real  $k_x$ and imaginary $q$,  the  (real) propagation constant $\beta_n$ must be  above the light line of the core ($\beta_n c< \omega n_1$)  and below the light line of the cladding ($\beta_n c> \omega n_0$)~\cite{Jannopoulos2008} [see \eqsref{wave-relations-a}{wave-relations-b}]. In contrast, leaky guided modes decay also inside the core, i.e., they have complex  $q$ and $k_x$. The propagation constants of the lowest-order leaky modes still sit above the light line of the core, but  at higher orders  or smaller frequencies,  we find modes below the light line, as demonstrated  in \figref{dispersion-fig}(a) when the red curves   penetrate the line $\mathrm{Re}[\beta_n] c= \omega n_1$.

\section{Scalar Maxwell equations as  Schr\"{o}dinger-type equations\label{transformation}} 

Apart from a very limited number of analytically solvable geometries, such as  the piecewise continuous geometry of our example system, it is generally impossible to construct simple  transcendental equations and one must solve \eqsref{TE}{TM} directly.  Since the TE  Maxwell equation is a Schr\"{o}dinger-type equation, it can be solved using standard approaches from quantum mechanics.  Although the TM equation contains a non-conservative force term [see discussion following \eqref{TM}], we can recast it as  a  Schr\"{o}dinger-type equation  by introducing the transformation: $h_y(x) = n(x)\psi(x)$. We find that the  new field $\psi$ satisfies \eqref{SE-well} with the  effective potential 
\begin{align}
V_\mathrm{TM}(x) = -\left(\tfrac{\omega}{c}\right)^2n^2(x) -\tfrac{1}{n}\tfrac{d^2n}{dx^2}  + 2\left(\tfrac{1}{n}\tfrac{dn}{dx}\right)^2
\label{eq:potential}
\end{align}
More generally, one can apply  similar tricks to transform the full-vector Maxwell equation into a  Schr\"{o}dinger-type equation, even in the absence of mirror-plane symmetry (for details, see lecture 3 in~\cite{berry1985classical}).

   \begin{figure*}[t]
     \includegraphics[scale=1]{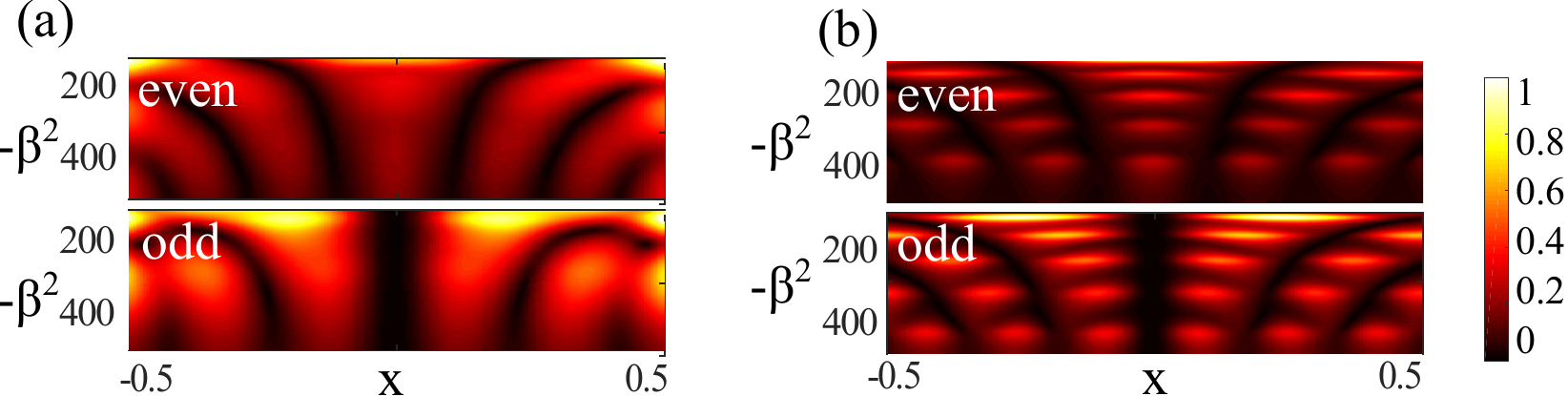}
     \caption{Normalized local density of states  [\eqref{LDOS}] for TE and TM modes [panels (a) and (b) respectively], for the structure from \figref{general-idea}. The local density of states vanishes (black regions) at nodes of the modes  and peaks at field maxima (yellow regions).  In the TM case, strong peaks are seen near resonant wave vectors. The color scale is shown on the right.
  }
    \label{fig:LDOS-fig}
\end{figure*}

The analogy to quantum mechanics offers a simple interpretation to the nature of the TE and TM solutions. Returning to our example system [\figref{general-idea}(a)], the index profile of the  rectangular waveguide   can be written as
\begin{align}
n(x) &= 
n_0 + (n_1 - n_0)\left[H(\tfrac{L}{2}+x) + H(\tfrac{L}{2}-x) -1\right],
\label{eq:sharp-n}
\end{align}
where $H(x)$ is the Heaviside step function.  The TE potential [\eqref{V-TE}] with $n(x)$ given by \eqref{sharp-n} is equivalent to a   one-dimensional square well. Confined modes   are analogous to bound states in quantum mechanics, and their real  propagation constants are in the range $-\omega^2n_1^2<-\beta^2_n<-\omega^2n_0^2$ (i.e., between the bottom of the well and the ``ionization threshold''), as shown in   \figref{dispersion-fig}(b). The effective TM potential [\eqref{potential}] with $n(x)$ given by \eqref{sharp-n}  is equivalent to a square-well potential with barriers of infinite height at the well boundaries. In order to visualize these barriers, we introduce  the smoothed index profile:
\begin{align}
&n_\alpha(x) =
n_0 + \tfrac{n_1 - n_0}{2}\left\{\tanh[\alpha(x+\tfrac{L}{2})] + \tanh[\alpha(x-\tfrac{L}{2})]\right\}
\label{eq:smooth-n}
\end{align}
which converges to $V_\mathrm{TM}$ in the limit of $\alpha\rightarrow\infty$. The TE and TM effective potentials, $V_\mathrm{TE}^{(\alpha)}$ anv $V_\mathrm{TM}^{(\alpha)}$ respectively,  with smoothing parameter $\alpha = 25$ are shown in \figref{dispersion-fig}(c). { The barriers in the TM potential give rise to   constructive interference of the scattered light and produce   a higher intensity  inside  the  waveguide  in  comparison  to  TE modes. This point is demonstrated in  \figref{dispersion-fig}(d), which shows  three even confined modes and  the first three leaky modes in the TE (blue solid lines) and TM (red dashed lines) polarizations.}

\section{Resonance structure in the  TE and TM   density of states\label{DOS-section}}

In non-Hermitian quantum mechanics, resonances are associated with peaks in the density of states.  In non-degenerate systems with weak loss or gain, the density of states is given by a sum over delta-function peaks at the bound-state energies and Lorentzian peaks at the resonant energies. In the context of non-Hermitian waveguides, the density of states is similarly defined as
\begin{align}
\rho(\beta) &=
\sum_{n } \delta(\beta^2 - [\beta_n^2]^\mathrm{con})+
\sum_n \mathrm{Im} \frac{1}{[\beta_n^2]^\mathrm{res} - \beta^2}.
\label{eq:DOS}
\end{align}
The first sum contains   confined modes and the second  contains the  leaky modes. The latter becomes a set of Lorentzian peaks in the limit of isolated resonances (i.e.,  when $\mathrm{Re}[\beta^2_n]\gg\mathrm{Im}[\beta^2_n]$), since in this limit
\begin{align}
\mathrm{Im} \frac{1}{\beta_n^2 - \beta^2} \approx
-\frac{\mathrm{Im}([\beta_n])/2\mathrm{Re}[\beta_n]}{(\beta - \mathrm{Re}[\beta_n])^2 +(\mathrm{Im}[\beta_n])^2 }.
\label{eq:DOS}
\end{align}
The density of states of  TE and TM leaky modes is shown in \figref{general-idea}(c) for the structure from panel (a). The modal structure is evident in the TM case and is absent in the TE spectrum.

When   the waveguide is excited at a specific location $(x_0,z_0)$   (instead of  homogeneously over the entire transverse cross-section), the system's response is determined by the local density of states, which is defined as\cite{Taflove2013} 
\begin{equation}
\rho_\mathrm{local}(x,\beta) = -\mathrm{Im}\left[\sum_n \frac{1}{\beta^2 - \beta_n^2} \frac{\psi^R_n(x)\psi^L_n(x)}{\int dx \psi_n^L(x)\psi^R_n(x)}\right].
\label{eq:LDOS}
\end{equation}
\begineqref{LDOS} includes both   leaky and confined modes in the summation and denotes  the right and left eigenvectors of   Maxwell operators  [\eqsref{TE}{TM}] by $\psi_n^R$ and $\psi_n^L$ respectively~\cite{Moiseyev2011}.  Since   Maxwell's equations have the form of  a symmetric generalized eigenvalue problem~\footnote{The symmetry of Maxwell's operator can be seen from its equivalence  to a symmetric Schrodinger-type equation.}, the  left and right eigenvectors are equal. In order to evaluate the denominator of \eqref{LDOS}, some care needs to be taken to handle the divergence of the leaky modes at $x=\pm\infty$. It turns out that the modes are properly normalized by omitting the outer limits of integration:
\begin{gather}
\int_{-\infty}^{\infty} \varepsilon(x)\psi_n^2(x) dx = \nonumber\\
\int^{-L/2}\varepsilon_0\psi_n^2(x) dx + \int_{-L/2}^{L/2} \varepsilon(x)\psi_n^2(x) dx + \int_{L/2} \varepsilon_0\psi_n(x) dx
\label{eq:unconjugated_norm}
\end{gather}
 (A rigorous proof of this normalization approach  can be found in~\cite{Moiseyev2011} and~\cite{pick2017general}.)
Substituting \eqref{ansatz} into \eqref{unconjugated_norm}, we obtain in our case
\begin{gather}
\int_{-\infty}^{\infty} dx \psi(x)^2 =\nonumber\\
 \frac{e^{-iqL}}{iq}+
\left(
A^2\frac{k_xL+\sin k_xL}{2k_x} + B^2\frac{k_xL-\sin k_xL}{2k_x} 
\right).
\label{eq:inner-product}
\end{gather}
\beginfigref{LDOS-fig} shows the normalized local density of states $\rho_\mathrm{local}(x,\beta)$ [\eqref{LDOS}] for TE and TM modes [panels (a) and (b) respectively], for the structure from \figref{general-idea}. The local density of states vanishes  at nodes of the f  (black regions) and culminates at field maxima (yellow regions).  In the TM case, strong peaks are seen near resonant wave vectors.

\section{Discussion\label{discussion-section}}

{ In this paper, we explored  the polarization dependence of the propagation distance  in \emph{perfectly straight real-index  waveguides}. We focused on a special kind of modes, in which the imaginary part of the  propagation constant is solely  due to leakage of radiation in the  transverse direction. Complex propagation constants are typically encountered in systems with a complex index of refraction, such as $\mathcal{PT}$-symmetric  waveguides~\cite{Guo2009} with commensurate amounts of loss and gain, and  in semiconductor lasers with nonlinear  gain~\cite{bogatov1998calculation}. They also arise in  waveguides with surface roughness or  waveguides with  small variation of the cross section along the waveguide axis~\cite{vorobeichik1995modal}. In  bent planar waveguides, the   bend losses can be described by assigning an   imaginary part to  the propagation constant~\cite{lu2005simple}.  In this context, recent work by Bauters \emph{et. al}  showed that  the TM  modes  in   rectangular   waveguides with a high aspect ratio  are associated with ultra-low  bend losses~\cite{bauters2011ultra,dai2010polarization}. This  property of TM modes in  bent waveguides is similar to our  findings in straight waveguides. }

 Since straight real-index waveguides are  much easier to fabricate than   the other mentioned examples, they  can be used to design simple experiments to test the predictions and applications of non-Hermitian optics.   For example, one can use leaky-mode propagation to design  simple and compact  filters for  TM-polarized light.   
{  While traditional TE/TM mode filters typically use composite structures, such as  metal-clad and buffer layers~\cite{ohke2002tm}, or anisotropic substrates~\cite{sosnowski1972polarization}, we propose  using straight single-constituent  waveguides.}  
Consider a     waveguide   whose width $L_x$ varies adiabatically  as a function of  $z$, so it consists of a wide  and a   narrow section. While the wide section supports $N$  confined  modes,  the narrow section supports only $N-1$ confined  modes.   Let us assume that the waveguide  initially  guides  unpolarized light, populating the $N$th  TE and TM  confined modes. Upon propagation, light enters the thin section and the  $N$th confined mode become leaky.     Since  the  TM leaky mode has a much longer propagation distance, the TE  mode  will decay to zero and only  the TM resonance  will survive after a relatively short propagation distance. This simple design     can be easily integrated on a microscale chip, since the thin section  can be  made very short assuming that  the contrast between the TE and TM propagation constants is significant.  Moreover,  similar principles  can be applied  to design a multimode filter.

Another intriguing application of TM leaky resonances is   communication between distant waveguides.  
 Confined modes can only carry information between nearby  waveguides.
 { 
  The separation between  the waveguides can not  exceed the length of the evanescent tails because the coupling strength  depends on the overlap  between  modes of the individual  waveguides~\cite{Haus1991}. (See, for instance,  Ref. \cite{ornigotti2008visualization}, which shows Rabi oscillations between evanescently  coupled waveguides.) By observing the leaky mode profiles in \figref{dispersion-fig}(d), we expect that leaky modes could convey information over many wavelengths of the light.  } TE resonances are not suitable for this task because the modes are delocalized  and only a small fraction of the light actually propagates in the core of the waveguide. However, TM resonances are promising candidates for  this task.

\section*{Acknowledgments}
A. P.  is partially supported  by an Aly Kaufman Fellowship at the Technion.
N. M.  acknowledges the financial support of I-Core: The Israeli Excellence Center ``Circle of Light,'' and of the Israel Science Foundation Grant No. 1530/15. Finally, 
we would like to thank Sir Michael V.  Berry, Steven G. Johnson, and Edvardas Narevicius for insightful discussions.

\end{document}